\documentclass[aps,prb,twocolumn,showpacs]{revtex4}
\usepackage{amssymb}

\usepackage{float}

\usepackage{graphicx}

\usepackage{tabularx}

\usepackage{color}

\begin{document}

\title{Superconductivity in the ternary compound SrPt$_{10}$P$_4$ with complex new structure}

\author{Bing Lv$^{1{\star}}$, BenMaan I. Jawdat$^2$, Zheng Wu$^2$, Sheng Li$^1$, and Ching-Wu Chu$^{3,4{\star}}$}

\affiliation{$^1$Department of Physics, University of Texas at Dallas,Richardson, TX 75080, USA}
\affiliation{$^2$Air Force Research Laboratory, Kirtland Air Force Base, Albuquerque, NM 87123, USA}
\affiliation{$^3$TcSUH and Department of Physics, University of Houston, TX 77204, USA}
\affiliation{$^4$Lawrence Berkeley National Laboratory, 1 Cyclotron Road, Berkeley, CA 94720, USA}

\begin{abstract}
We report superconductivity at 1.4K in the ternary SrPt$_{10}$P$_4$ with a complex new structure. SrPt$_{10}$P$_4$ crystallizes in a monoclinic space-group C2/c (\#15) with lattice parameters a= 22.9151(9)${\AA}$, b= 13.1664(5)${\AA}$, c=13.4131(5)${\AA}$, and $\beta$= 90.0270(5)${^\circ}$. Bulk superconductivity in the samples has been demonstrated through resistivity, ac susceptibility, and heat capacity measurements. High pressure measurements have shown that the superconducting T$_C$ is systematically suppressed upon application of pressure, with a dT$_C$/dP coefficient of -0.016 K/GPa.
\end{abstract}

 \maketitle

\section{Introduction}
The discovery of Fe-pnictide superconductors with T$_C$ up to 57K\cite{1, 2} has stimulated many research efforts to search for new superconductors in other transition metal pnictide compounds and in their Fe-based layered analogs with related or unrelated structures. Among different types of Fe-pnictide discovered to date, the so-called "122" materials\cite{3,4,5} with ThCr$_2$Si$_2$-type structures are of particular interest. More than 700 compounds of AM$_2$X$_2$ adopt this structures\cite{6}, and a variety of interesting physics, such as valence fluctuations\cite{7, 8}, heavy fermion behavior\cite{9, 10}, and magnetic properties\cite{11, 12, 13}, have been discussed before.  Questions, therefore, are raised that whether superconductivity could be found in other materials which adopts similar type structures. In fact, many new superconducting compounds with related structures have been discovered, including BaNi$_2$P$_2$ (T$_C$=3K)\cite{14}, BaNi$_2$As$_2$ (T$_C$= 0.7K)\cite{15} and BaIr$_2$As$_2$ (T$_C$=2.45K)\cite{16}, with a ThCr$_2$Si$_2$-type structure, SrPt$_2$As$_2$ (T$_C$=5.2 K)\cite{17} and BaPt$_2$Sb$_2$ (T$_C$=1.8K)\cite{18}with a CaBe$_2$Ge$_2$-type structure, and SrPtAs (T$_C$=2.4K) with a MgB$_2$-type structure\cite{19}. On the other hand, many Pt-based superconductors with unrelated structures have been discovered which include noncentrosymmetric BaPtSi$_3$\cite{20}, Li$_2$Pt$_3$B\cite{21, 22}, heavy Fermion CePt$_3$Si\cite{23}, and the recently discovered centrosymmetric SrPt$_3$P (T$_C$=8.4 K)\cite{24}. SrPt$_3$P crystallizes in an antiperovskite structure that is very similar to that of CePt$_3$Si but displays centrosymmetry. Further calorimetric studies suggested that it is a conventional strong electron-phonon coupling superconductor, but further theoretical calculations have brought a new perspective\cite{25, 26}, i.e. one may be able to tune the structure from centrosymmetric to noncentrosymmetric through chemical doping or application of pressure\cite{25} and to induce possible unconventional superconductivity in this compound. Various experimental efforts\cite{27, 28, 29} have been carried out, but unfortunately have not shown significant change of the structures. In the course of chemical doping studies of SrPt$_3$P, we have discovered several new compounds, namely SrPt$_6$P$_2$ (T$_C$=0.6K)\cite{30} and SrPt$_{10}$P$_4$. SrPt$_{10}$P$_4$ crystallizes in a very complex structure (with a total of 240 atoms in one unit cell). Resistivity, ac susceptibility, and heat capacity measurements have demonstrated bulk superconductivity at 1.4K in this new compound. High pressure measurements have shown that the superconducting T$_c$ is systematically suppressed upon application of pressure, with a dT$_C$/dP coefficient of -0.016 K/GPa.
\begin{figure}[H]
\includegraphics[width=8.5cm, bb=10 95 190 240]{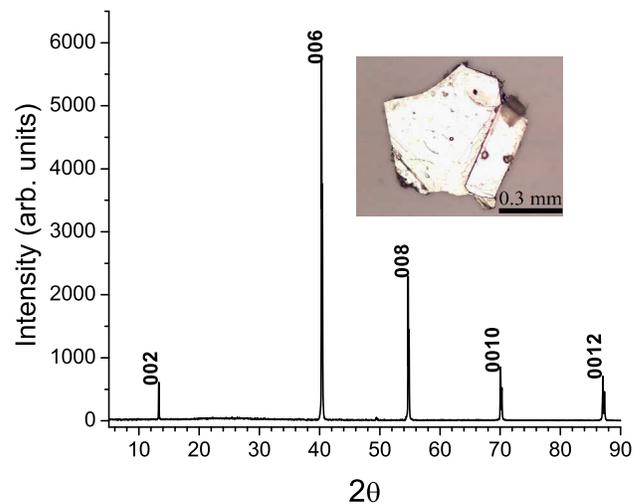}
\caption {(color online) X-ray diffraction pattern on a selected crystal of SrPt$_{10}$P$_4$. Inset: the crystal with size $\sim$ 0.7mm.} \label{fig1}
\end{figure}

\begin{table*}[t]
\centering
\caption {Crystal structure of SrPt$_{10}$P$_4$ at room temperature}
\begin{tabularx}{\textwidth}{*{5}{>{\centering\arraybackslash}X}}
\hline
\multicolumn{5}{c}{Crystal system: Monoclinic \qquad Space group: C2/c(No. 15) \qquad Total reflection: 12841}\\
\multicolumn{5}{c}{Independent reflection: 4963 \qquad Absorption coefficient: 143.456 mm$^{-1}$ \qquad Goodness-of-fit: 1.003 }
\\
\multicolumn{5}{c}{Unit cell parameter: a=22.9151(9)${\AA}$, b=13.1664(5)${\AA}$, c=13.4131(5)${\AA}$, $\beta$=90.0270${^\circ}$}\\
\hline
\multicolumn{5}{c}{Atomic position}\\
\hline
\hline
atom & Wyckoff & x/a & y/b & z/c\\
\hline
\multicolumn{1}{c}{Pt1}&\multicolumn{1}{c}{8f}&\multicolumn{1}{c}{0.0782(1)}&\multicolumn{1}{c}{0.2746(1)}& \multicolumn{1}{c}{0.0164(1)}\\
\multicolumn{1}{c}{Pt2}&\multicolumn{1}{c}{8f}&\multicolumn{1}{c}{0.3247(1)}&\multicolumn{1}{c}{-0.0210(1)}&\multicolumn{1}{c}{0.5158(1)}\\
\multicolumn{1}{c}{Pt3}&\multicolumn{1}{c}{8f}&\multicolumn{1}{c}{0.2477(1)}&\multicolumn{1}{c}{-0.0144(1)}&\multicolumn{1}{c}{0.3512(1)}\\
\multicolumn{1}{c}{Pt4}&\multicolumn{1}{c}{8f}&\multicolumn{1}{c}{0.1212(1)}&\multicolumn{1}{c}{0.1487(1)}&\multicolumn{1}{c}{0.3431(1)}\\
\multicolumn{1}{c}{Pt5}&\multicolumn{1}{c}{8f}&\multicolumn{1}{c}{0.1513(1)}&\multicolumn{1}{c}{0.0028(1)}&\multicolumn{1}{c}{0.4813(1)}\\
\multicolumn{1}{c}{Pt6}&\multicolumn{1}{c}{8f}&\multicolumn{1}{c}{0.1178(1)}&\multicolumn{1}{c}{0.1182(1)}&\multicolumn{1}{c}{0.1438(1)}\\
\multicolumn{1}{c}{Pt7}&\multicolumn{1}{c}{8f}&\multicolumn{1}{c}{0.1267(1)}&\multicolumn{1}{c}{0.1085(1)}&\multicolumn{1}{c}{0.6543(1)}\\
\multicolumn{1}{c}{Pt8}&\multicolumn{1}{c}{8f}&\multicolumn{1}{c}{0.0016(1)}&\multicolumn{1}{c}{0.2656(1)}&\multicolumn{1}{c}{-0.1478(1)}\\
\multicolumn{1}{c}{Pt9}&\multicolumn{1}{c}{8f}&\multicolumn{1}{c}{0.0956(1)}&\multicolumn{1}{c}{0.2515(1)}&\multicolumn{1}{c}{0.5156(1)}\\
\multicolumn{1}{c}{Pt10}&\multicolumn{1}{c}{8f}&\multicolumn{1}{c}{0.1183(1)}&\multicolumn{1}{c}{0.3654(1)}&\multicolumn{1}{c}{-0.1558(1)}\\
\multicolumn{1}{c}{Pt11}&\multicolumn{1}{c}{8f}&\multicolumn{1}{c}{0.2321(1)}&\multicolumn{1}{c}{-0.0210(1)}&\multicolumn{1}{c}{0.6514(1)}\\
\multicolumn{1}{c}{Pt12}&\multicolumn{1}{c}{8f}&\multicolumn{1}{c}{0.3538(1)}&\multicolumn{1}{c}{0.1168(1)}&\multicolumn{1}{c}{0.3520(1)}\\
\multicolumn{1}{c}{Pt13}&\multicolumn{1}{c}{8f}&\multicolumn{1}{c}{0.1925(1)}&\multicolumn{1}{c}{0.2650(1)}&\multicolumn{1}{c}{-0.0049(1)}\\
\multicolumn{1}{c}{Pt14}&\multicolumn{1}{c}{8f}&\multicolumn{1}{c}{0.1344(1)}&\multicolumn{1}{c}{0.3937(1)}&\multicolumn{1}{c}{0.1549(1)}\\
\multicolumn{1}{c}{Pt15}&\multicolumn{1}{c}{8f}&\multicolumn{1}{c}{0.3675(1)}&\multicolumn{1}{c}{0.1309(1)}&\multicolumn{1}{c}{0.6453(1)}\\
\multicolumn{1}{c}{Pt16}&\multicolumn{1}{c}{8f}&\multicolumn{1}{c}{0.0356(1)}&\multicolumn{1}{c}{0.0800(1)}&\multicolumn{1}{c}{0.4892(1)}\\
\multicolumn{1}{c}{Pt17}&\multicolumn{1}{c}{8f}&\multicolumn{1}{c}{0.0585(1)}&\multicolumn{1}{c}{0.4887(1)}&\multicolumn{1}{c}{-0.0014(1)}\\
\multicolumn{1}{c}{Pt18}&\multicolumn{1}{c}{8f}&\multicolumn{1}{c}{0.1037(1)}&\multicolumn{1}{c}{0.1359(1)}&\multicolumn{1}{c}{-0.1493(1)}\\
\multicolumn{1}{c}{Pt19}&\multicolumn{1}{c}{8f}&\multicolumn{1}{c}{0.2134(1)}&\multicolumn{1}{c}{0.1710(1)}&\multicolumn{1}{c}{0.5131(1)}\\
\multicolumn{1}{c}{Pt20}&\multicolumn{1}{c}{8f}&\multicolumn{1}{c}{-0.0129(1)}&\multicolumn{1}{c}{0.2723(1)}&\multicolumn{1}{c}{0.1508(1)}\\
\multicolumn{1}{c}{Sr1}&\multicolumn{1}{c}{4e}&\multicolumn{1}{c}{0}&\multicolumn{1}{c}{0.0089(2)}&\multicolumn{1}{c}{0.25}\\
\multicolumn{1}{c}{Sr2}&\multicolumn{1}{c}{8f}&\multicolumn{1}{c}{0.2495(1)}&\multicolumn{1}{c}{0.2549(2)}&\multicolumn{1}{c}{0.2468(2)}\\
\multicolumn{1}{c}{Sr3}&\multicolumn{1}{c}{4e}&\multicolumn{1}{c}{0}&\multicolumn{1}{c}{0.4958(2)}&\multicolumn{1}{c}{-0.25}\\
\multicolumn{1}{c}{P1}&\multicolumn{1}{c}{8f}&\multicolumn{1}{c}{0.3122(2)}&\multicolumn{1}{c}{0.1592(4)}& \multicolumn{1}{c}{0.5023(5)}\\
\multicolumn{1}{c}{P2}&\multicolumn{1}{c}{8f}&\multicolumn{1}{c}{0.2337(2)}&\multicolumn{1}{c}{-0.1085(4)}&\multicolumn{1}{c}{0.5007(4)}\\
\multicolumn{1}{c}{P3}&\multicolumn{1}{c}{8f}&\multicolumn{1}{c}{0.0620(2)}&\multicolumn{1}{c}{0.0934(4)}&\multicolumn{1}{c}{0.0016(4)}\\
\multicolumn{1}{c}{P4}&\multicolumn{1}{c}{8f}&\multicolumn{1}{c}{0.0829(2)}&\multicolumn{1}{c}{0.2629(4)}&\multicolumn{1}{c}{0.2209(5)}\\
\multicolumn{1}{c}{P5}&\multicolumn{1}{c}{8f}&\multicolumn{1}{c}{-0.0874(2)}&\multicolumn{1}{c}{0.2561(4)}&\multicolumn{1}{c}{-0.2202(5)}\\
\multicolumn{1}{c}{P6}&\multicolumn{1}{c}{8f}&\multicolumn{1}{c}{0.1569(2)}&\multicolumn{1}{c}{0.0014(4)}&\multicolumn{1}{c}{0.2791(5)}\\
\multicolumn{1}{c}{P7}&\multicolumn{1}{c}{8f}&\multicolumn{1}{c}{0.3242(2)}&\multicolumn{1}{c}{-0.0064(4)}&\multicolumn{1}{c}{0.7227(5)}\\
\multicolumn{1}{c}{P8}&\multicolumn{1}{c}{8f}&\multicolumn{1}{c}{0.0130(2)}&\multicolumn{1}{c}{0.6405(4)}&\multicolumn{1}{c}{0.0009(5)}\\
\hline
\end{tabularx}

\end{table*}

\begin{figure*}[t]
\includegraphics[width=17cm, bb=0 0 490 195]{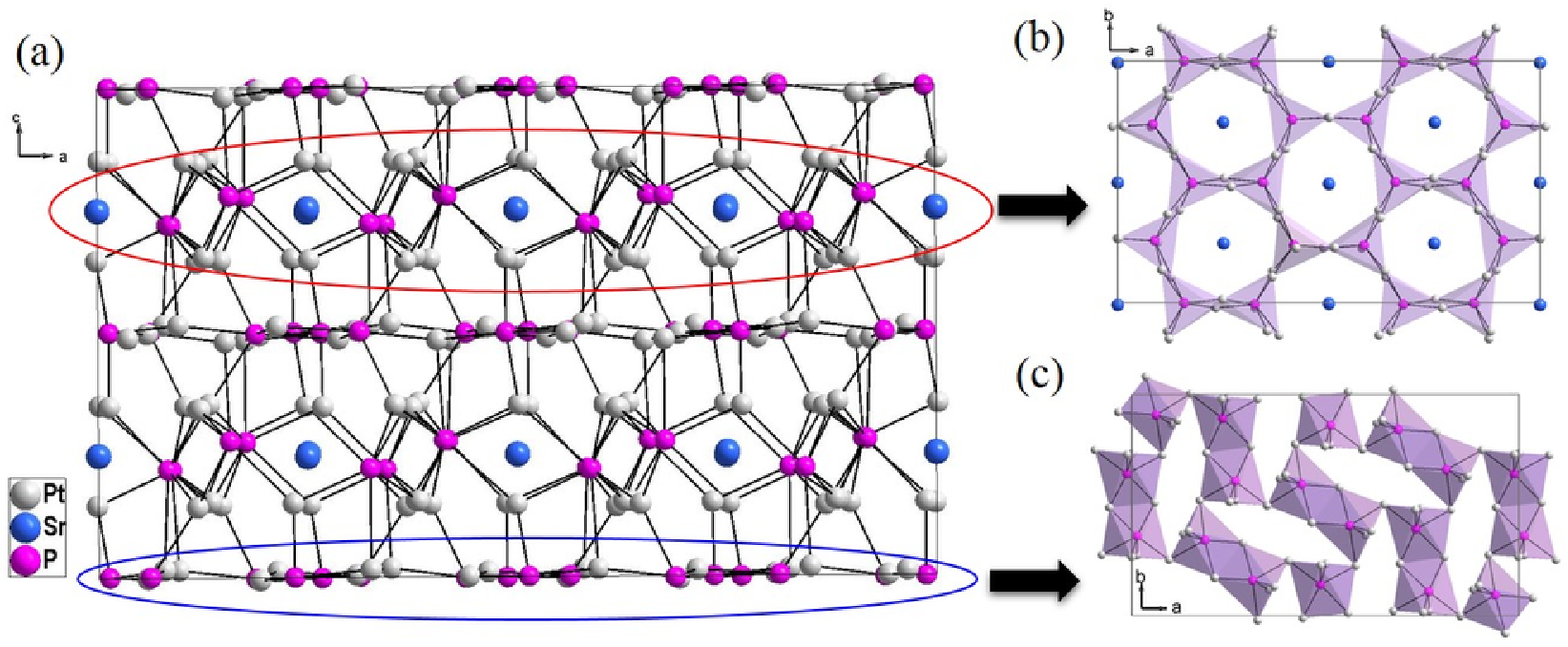}
\caption {(color online) Crystal structure of SrPt$_{10}$P$_4$:  a) Unit cell view along b axis showing a complex three-dimensional network, with two distinct layers highlighted by ellipses; b) polyhedra top view of one layer along c axis with distorted trigonal prismatic Pt$_6$P units forming a honeycomb-like network;  c) polyhedra top view of another layer along c axis that consists of highly distorted Pt$_6$P octahedral building blocks.} \label{fig2}
\end{figure*}

\section{Experimental section}

The polycrystalline samples were prepared by high temperature reactions of stoichiometric Sr pieces (Alfa Aesar, 99.95$\%$), Pt powder
(Alfa Aesar, $>$99.95$\%$), and prereacted PtP$_2$ from Pt powder and P powder (Sigma Aldrich, $>$99.99$\%$) within an alumina crucible that was sealed inside a clean and dried quartz tube under vacuum.  The tube was placed in a furnace, heated slowly up to 1000${^\circ}$C overnight, and maintained at 1000${^\circ}$C for four days before being slowly cooled down to 400${^\circ}$C with a rate of 0.5${^\circ}$C/min. The assembly was finally quenched in ice water from 400${^\circ}$C to avoid the possible formation of white phosphorus caused by any unreacted P.  To improve the homogeneity, the sample was reground, pelletized, and reheated following the previously described temperature profile. All synthesis procedures were carried out within a purified Ar-atmosphere glovebox with total O$_2$ and H$_2$O levels $<$0.1 ppm. Single crystal XRD data was collected using a Bruker SMART APEX diffractometer equipped with 1K CCD area detector using graphite-monochromated Mo K$_\alpha$ radiation. The electrical resistivity $\rho$(T, H) was measured by employing a standard 4-probe method, and heat capacity data were collected using a relaxation method down to 0.5 K under magnetic field up to 1 T using a $_3$He-attachment in a Quantum Design Physical Property Measurement System.  The ac magnetic susceptibility at 15.9 Hz as a function of temperature, $\chi$(T), was measured by employing a compensated dual coil for mutual inductance measurement using the Linear Research LR 400 Bridge.  High-pressure resistivity measurements up to 18 kbar were conducted using a BeCu piston-cylinder cell with Fluorinert77 as the quasihydrostatic pressure medium. A lead manometer was used to measure the pressure in situ with the LR 400 Inductance Bridge\cite{31, 32}.

\section{Results and Discussion}

The as-synthesized pellet has a dark grey color, and is stable in air. On its top, many small shiny crystals with metallic luster are found, indicating that the material is a congruently melting compound, and that larger size crystals could be obtained through slow cooling process from melt. Indeed, we obtained large crystals with size up to 0.7 mm (as shown in the inset to Fig. 1), which were subsequently used for electrical resistivity and magnetic susceptibility measurements.  X-ray diffraction studies reveal a c-axis preferred orientation, as shown in Fig. 1. The (004) peak is too weak to be observed, consistent with the theoretical pattern generated from crystal structures determined by X-ray single-crystal diffraction. Small blocks of shiny single crystals with a typical size of 0.04 x 0.04 x 0.06 mm$^3$ were isolated for single-crystal X-ray diffraction to determine the crystal structure. The refined composition is SrPt$_{10}$P$_4$, which is consistent with SEM-WDS results of Sr:Pt:P=1:9.66(1):3.67(2). The compound crystallizes in a monoclinic space-group C2/c (\#15) with lattice parameters a= 22.9151(9)${\AA}$, b= 13.1664(5)${\AA}$, c=13.4131(5)${\AA}$, $\beta$= 90.0270(5)${^\circ}$ and z=16. It should be noted that we have carried out symmetry tests on all of the refined atom positions by using the program PLATON and concluded that C2/c is the correct space group and that no higher symmetry can be found. Arbitrarily merging data to the orthorhombic Laue group mmm results in an unusually high R(int) = 0.427, as compared to the actual monoclinic Laue group 2/m with R(int) = 0.039. This further confirms that the compound indeed crystallizes in a monoclinic, not orthorhombic, structure. The detailed crystallographic information, including the Wyckoff position of individual atoms, is listed in Table I.

The structure of SrPt$_{10}$P$_4$ is rather complicated with its unit cell shown in Fig. 2a. The fundamental build unit is a 6-coordinated P-centered highly distorted Pt$_6$P octahedral or trigonal prism, as seen previously in the SrPt$_3$P\cite{11} and SrPt$_6$P$_2$\cite{15} structures.  The crystal structure could be considered as two distinct types of layers intertwined with one another, shown in Fig. 2a (projection view along b axis), forming a complex three-dimensional structure. One layer consists of distorted trigonal prismatic Pt$_6$P units that are edge-shared with each other and form a honeycomb-like network as seen in Fig. 2b (projection view along c axis). Sr atoms occupy the centers of the honeycomb-like network (crystallographic 4e site).  The other layer is composed of a network of highly distorted Pt$_6$P octahedral building blocks. These Pt$_6$P octahedra are edge-shared first and forms pairs. These pairs of edge-shared octahedra are then corner shared with one another and form a network in which the Pt-Pt distance of neighboring Pt$_6$P octahedra is too short ($\sim$2.668${\AA}$) to accommodate any Sr atom, as seen in Fig. 2c (projection view along c axis). 

The distorted octahedral Pt$_6$P is reminiscent of the distorted anti-perovskite Pt$_6$P building blocks in SrPt$_3$P.  In SrPt$_3$P, the Pt$_6$P octahedra are corner-shared, arranged antipolar, and thus formed stoichiometry as "Pt$_{6/2}$P=Pt$_3$P". But in StPt$_{10}$P$_4$, these Pt$_6$P octahedra are arranged completely differently, where they are edge-shared forming ``Pt$_{10}$P$_2$", and then corner-shared. To facilitate such an arrangement, the Pt$_6$P octahedr is much more distorted than those in SrPt$_3$P. Similarly, the trigonal prismatic Pt6P building blocks in SrPt$_{10}$P$_4$ are very similar to those found in SrPt$_6$P$_2$, but less distorted to accommodate the edge-sharing coordination in SrPt$_{10}$P$_4$, rather than corner-shared feature in SrPt$_6$P$_2$.

The electrical resistivity $\rho$(T) of SrPt$_{10}$P$_4$ from 270K down to 0.5K is shown in Fig. 3. The room temperature resistivity of SrPt$_{10}$P$_4$ is $\sim$ 300$\mu\Omega\cdot$cm, smaller than that of SrPt$_3$P and SrPt$_6$P$_2$, and consistent with the expected trend toward increased metallicity with increased Pt:P ratio in the compound. The temperature dependence of the resistivity has a typical metallic behavior, with a much stronger negative curvature in the normal state than is seen in SrPt$_3$P and SrPt$_6$P$_2$, indicating the stronger electron correlations in this material. The relatively high value of residual resistivity ratio (RRR), $\rho$(270 K)/$\rho$(1.5 K)=32, suggests that the sample is of high quality. The resistivity drops sharply to zero below 1.4 K at zero field, characteristic of a superconducting transition.   The narrow width of the superconducting transition (less than 0.1 K) indicates the high quality of the sample. In the presence of a magnetic field, the superconducting transition is systematically broadened and shifted to lower temperature, and is suppressed to below 0.5K upon the application of a magnetic field of 1T, as shown in the inset to Fig. 3.

\begin{figure}[H]
\includegraphics[width=8.5cm, bb=15 90 220 235]{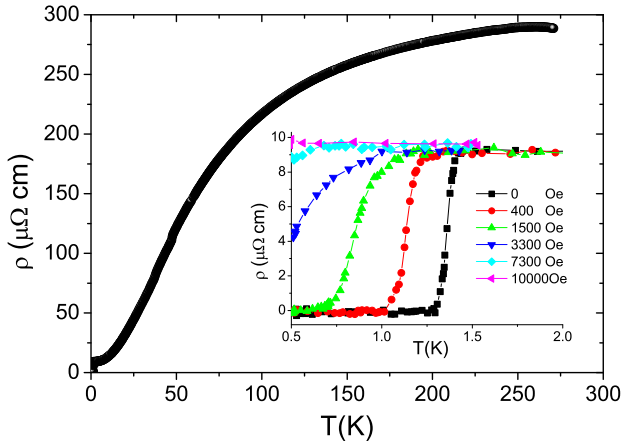}
\caption {(color online) Resistivity data of SrPt$_{10}$P$_4$ at H = 0 between 270K and 0.5 K. Inset: Resistivity data of SrPt$_{10}$P$_4$ under different magnetic field between 2.0 K and 0.5 K.} \label{fig3}
\end{figure}

\begin{figure}[H]
\includegraphics[width=8.5cm, bb=15 0 340 270]{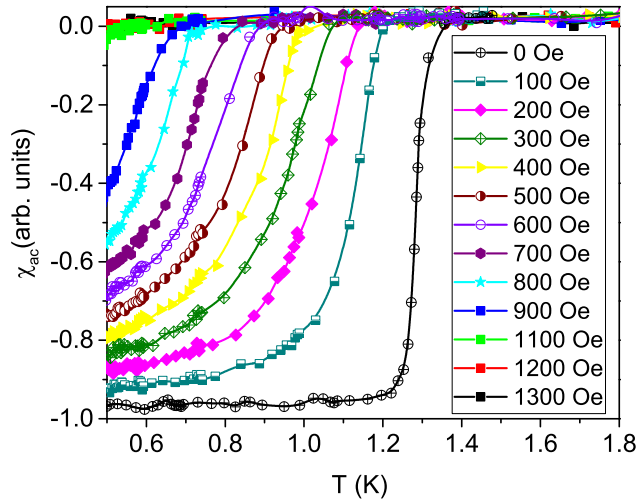}
\caption {(color online) ac magnetic susceptibility vs. temperature for SrPt$_{10}$P$_4$ at different applied magnetic fields between 1.8K and 0.5 K.} \label{fig4}
\end{figure}
Fig. 4 shows the ac susceptibility data of StPt$_{10}$P$_4$, where a large and narrow diamagnetic shift starting from 1.4K is clearly visible. This magnetic susceptibility is shifted toward the left upon applying magnetic field as expected, and also reveals the suppression of T$_C$ with increasing field.

In order to further verify the bulk nature of the superconductivity in SrPt$_{10}$P$_4$, we carried out specific heat measurements.  Because the mass of the largest crystal we have is $\sim$0.1mg, which falls below the critical mass needed for the specific heat measurement, we decided to measure a pure, bulk, polycrystalline sample. Figure 5 (a) shows the raw data of this measurement under different applied magnetic fields up to 4000 Oe. The jump at T$_C$ observed in the specific heat data clearly demonstrates the bulk nature of the superconductivity in the sample.

\begin{figure}[H]
\includegraphics[width=8.5cm, bb=5 120 165 350]{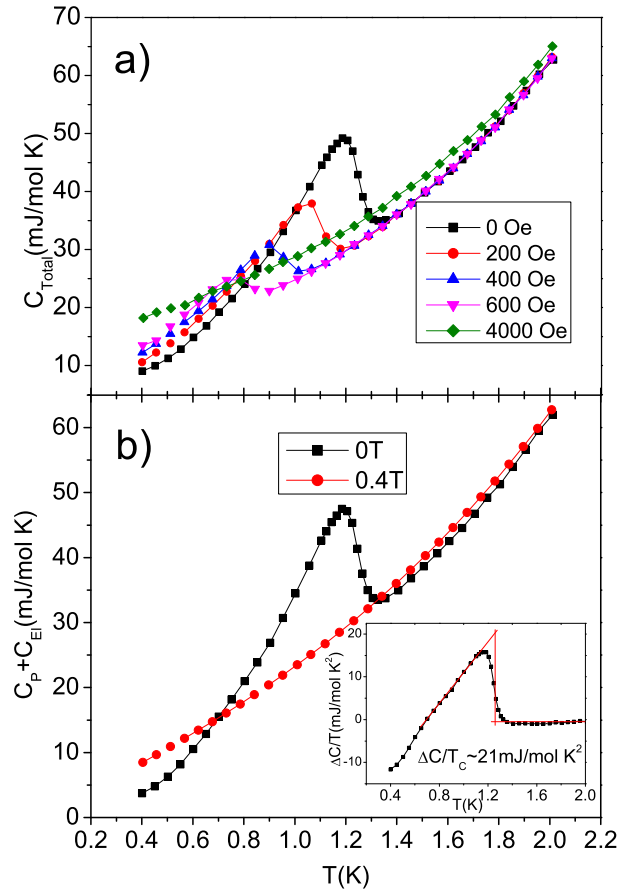}
\caption {(color online) Specific heat data: a) the raw data of specific heat under different magnetic field up to 0.4T, and b) specific data of normal and superconducting states through subtracting the Schottky anomaly. Inset: the difference of electronic specific heat between the superconducting and normal states.}
\label{fig5}
\end{figure}

It can also be clearly seen that the superconducting and normal state specific heat data deviate from one another, which may be caused by the Schottky anomaly at low temperature range. Using C$_{Total}$=C$_{El}$+C$_P$+C$_{Sch}$=$\gamma_N$T+$\beta$T$^3$ + nR($\frac{\Delta}{k_BT}$)$^2$$\exp$($\frac{\Delta}{k_BT}$)/(1+$\exp$($\frac{\Delta}{k_BT}$))$^2$, where $\Delta$ is the energy between two levels by considering spin  J=1/2\cite{33}, we can obtain the contributions of electrons (C$_{El}$) and phonons (C$_P$) to the total specific heat (C$_{Total}$) by subtracting that associated with the Schottky anomaly (C$_{Sch}$). The normal state electron and phonon contribution (Fig. 5b, line with red circles) was obtained by subtracting C$_{Sch}$(0.4T) from the normal state C$_{Total}$(T) at 0.4T. By fitting C$_{Total}$(T, 0.4T) through the Debye model, we obtained a Sommerfeld coefficient $\gamma_N$ = 20.6 mJ/mol K$^2$ and $\beta$ = 2.64 mJ/mol K$^4$, which correspond to the electronic and lattice contributions to the specific heat, respectively.  The Debye temperature can be deduced from the $\beta$ value through the relationship $\Theta_D$= (12$\pi^4$k$_B$N$_A$Z/5$\beta$)$^{1/3}$ and the obtained Debye temperature $\Theta_D$= 223K. Through a similar approach, we can also obtain the zero field electron and phonon specific heat (Fig. 5b, line with black squares) by subtracting the zero field Schottky contribution C$_{Sch}$(T, 0T). By subtracting the normal state specific heat from the superconducting one, we can get $\Delta$C$_{El}$/T$_C$ $\sim$ 21 mJ/mol K$^2$, which yields a $\Delta$C$_{El}$/$\gamma_N$T$_C$ of about 1.02, as seen in the inset to Fig. 5b. This value is comparable to the 1.2 for SrPt$_6$P$_2$\cite{34} but much smaller than the 2 for SrPt$_3$P. Such a small value of $\Delta$C$_{El}$/$\gamma_N$T$_C$ indicates weak coupling in this compound in comparison with SrPt$_3$P. Our preliminary specific heat fitting and critical field analysis strongly suggest multiple gap superconductivity feature for this compound, which will be addressed separately.

\begin{figure}[H]
\includegraphics[width=8.5cm, bb=0 0 380 270]{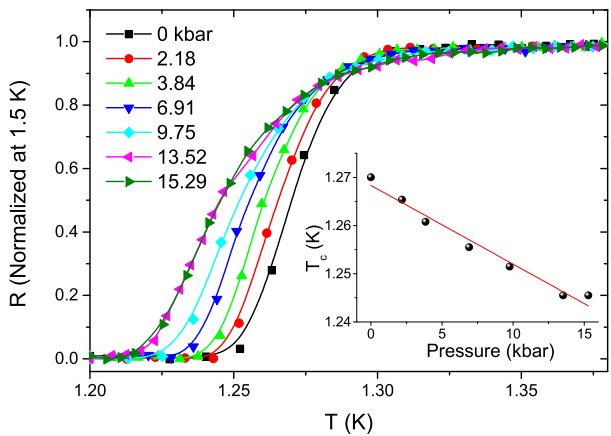}
\caption {(color online)  The normalized resistivity vs. temperature at different applied pressure. Inset: the pressure-dependent T$_C$ plot.} \label{fig6}
\end{figure}

The Kadowaki-Woods ratio R$_{KW}$=A/$\gamma_N^2$ has been used to judge the correlation of a metal where A is the quadratic term of resistivity of a Fermi liquid, and $\gamma_N$ is the Sommerfeld coefficient of the specific heat\cite{35}. The R$_{KW}$ is found to be a constant value for transition metals($\sim10^{-6}\mu\Omega cm K^2 mol^2/mJ^2$) and heavy fermion compounds($\sim10^{-5}\mu\Omega cm K^2 mol^2/mJ^2$)\cite{36}. We, therefore, have fitted the low temperature resistivity data using $\rho=\rho_0+AT^2$ and get A = 0.032$\mu\Omega$ cm /K$^2$. Combined with the $\gamma_N$ value obtained above, the calculated R$_{KW}$ is 7.5*10$^{-5}$$\mu\Omega$ cm K$^2$ mol$^2$/mJ$^2$, which indicates strong correlation in this material.

To probe the effects of high pressure on SrPt$_{10}$P$_4$, we applied high physical pressure using a BeCu piston-cylinder type pressure cell. Fig. 6 shows the resistivity data of SrPt$_{10}$P$_4$ under different applied pressure.  The superconducting transition is slightly lower and is relatively broader with the onset of the resistivity drop at 1.33K and zero resistance at $\sim$1.25 K. The broader superconducting transition may be related to the polycrystalline nature of the samples used for this measurement (as well as possible sample degradation
during the preparation) and grain-grain coherence within the sample.  A systematic suppression of the transition temperature with increasing pressure is
clearly visible. Taking a 50$\%$ drop of resistivity as the criteria of T$_C$, one can obtain a linear fit of T$_C$ vs. pressure, shown in the inset to Fig. 6,
which yields a pressure coefficient dT$_C$/dP = -0.016 K /GPa. The systematic suppression of T$_C$ suggests that there is no significant peak in the density of states near the Fermi level. The relatively small change in T$_C$ with pressure is comparable to that of many elemental superconductors, which exhibit a linear suppression of T$_C$ with pressure near ambient, and is very close to the value of dT$_C$/dP = -0.02 K/GPa for pure niobium metal\cite{37}. The suppression of T$_C$ with pressure in SrPt$_{10}$P$_4$ can therefore be explained as the result of a stiffening of the lattice induced by the pressure, which results in a weakening of the electron-phonon coupling or slightly decreasing of the density of state(DOS) at the Fermi surface.

In summary, a new ternary compound SrPt$_{10}$P$_4$ with a new structure type is synthesized through high-temperature solid state reactions, and its crystal structure is determined by X-ray single crystal diffraction. The compound crystallizes in a very complex three-dimensional structure that consists of two distinct layers based on P-centered highly distorted Pt$_6$P octahedral or trigonal prismatic building units. We have carried out systematic magnetization, electrical resistivity, and specific heat measurements, and demonstrated the bulk superconductivity with T$_C$ at 1.4K in this material. High pressure measurements have shown that the superconducting T$_C$ is systematically suppressed upon applying pressure, with a dT$_C$/dP coefficient of -0.016 K/GPa.

\begin{acknowledgments}
The authors would like to thank X. Q. Wang for the help with single crystal diffraction measurement. This work in Houston is supported in part by US Air Force Office of Scientific Research Grant No. FA9550-15-1-0236, the T.L.L. Temple Foundation, the John J. and Rebecca
Moores Endowment, and the State of Texas through the Texas Center for Superconductivity at the University of Houston.  B. Lv and S. Li also would like to acknowledge the funding support from US Air Force Office of Scientific Research and the Start-Up funds from University of Texas at Dallas.

\end{acknowledgments}

\noindent $^{\star}$ To whom correspondence should be addressed:

\noindent blv@utdallas.edu, cwchu@uh.edu


\begin{thebibliography}{99}
\bibitem{1} Y. Kamihara, T. Watanabe, M. Hirano, and H. Hosono, J. Am. Chem. Soc., {\bf130}, 3296 (2008).
\bibitem{2} Z. Wei, H. Li, W. L. Hong, Z. Lv, H. Wu, X. Guo, and K. Ruan, J. Supercond. Nov. Magn. {\bf21}, 213(2008).
\bibitem{3} M. Rotter, M. Tegel, and D. Johrendt, Phys. Rev. Lett. {\bf101}, 107006 (2008).
\bibitem{4} K. Sasmal, B. Lv, B. Lorenz, A. M Guloy, F. Chen, Y. Y. Xue,  and C. W. Chu, Phys. Rev. Lett., {\bf101}, 107007(2008).
\bibitem{5}	N. Ni, S. Nandi, A. Kreyssig, A. I. Goldman, E. D. Mun, S. L. Bud'ko, and P. C. Canfield, Phys. Rev. B {\bf78}, 014523(2008).

\bibitem{6} R. Hoffman, and C. Zheng, J. Phys. Chem., {\bf89}, 4175(1985).
\bibitem{7} E. R. Bauminger, D. Froindlich, I. Nowik, S. Ofer, I. Felner, and I. Mayer, Phys. Rev. Lett. {\bf30}, 1053 (1973).
\bibitem{8}	J. R. Neilson, T. M. McQueen, A. Llobet, J. Wen, and M. R. Suchomel, Phys. Rev. B {\bf87}, 045124 (2013).
\bibitem{9}	F. Steglich, J. Aarts, C. D. Bredl, W. Lieke, D. Meschede, W. Franz, and H. Sch\"{a}fer, Phys. Rev. Lett. {\bf43}, 1892(1979).
\bibitem{10} Z. Hossain, C. Geibel, F. Weickert, T. Radu, Y. Tokiwa, H. Jeevan, P. Gegenwart, and F. Steglich, Phys. Rev. B {\bf72}, 094411 (2005).
\bibitem{11} U. B. Paramanik, R. Prasad, C. Geibel, Z. Hossain, Phys. Rev. B {\bf89}, 144423 (2014).
\bibitem{12} V. K. Anand and D. C. Johnston, Phys. Rev. B {\bf91}, 184403 (2015).
\bibitem{13} M. Imai, C. Michioka, H. Ueda, and K. Yoshimura, Phys. Rev. B {\bf91}, 184414 (2015).
\bibitem{14} T. Mine, H. Yanagi, T. Kamiya, Y. Kamihara, M. Hirano, and H. Hosono. Solid State Commun. {\bf147}, 111(2008).
\bibitem{15} F. Ronning, N. Kurita, E. D. Bauer, B. L. Scott, T. Park, T. Klimczuk, R. Movshovich, and J. D. Thompson: J. Phys.: Condens. Matter {\bf20}, 342203(2008).
\bibitem{16} X. C. Wang, B. B. Ruan, J. Yu, B. J. Pan, Q. G. Mu, T.Liu, G. F. Chen, Z. A. Ren, Supercond. Sci. Technol., {\bf30}, 035007(2017).
\bibitem{17} K. Kudo, Y. Nishikubo, and M. Nohara: J. Phys. Soc. Jpn. {\bf79}, 123710(2010).


\bibitem{18} M. Imai, S. Ibuka, N. Kikugawa, T. Terashima, S. Uji, T. Yajima, H. Kageyama, and I. Hase Phys. Rev. B {\bf91}, 014513(2015).
\bibitem{19} Y. Nishikubo, K. Kudo, M. Nohara, J. Phys. Soc. Jpn. {\bf80}, 055002 (2011).
\bibitem{20} E. Bauer, R. T. Khan, H. Michor, E. Royanian, A. Grytsiv, N. Melnychenko-Koblyuk, P. Rogl, D. Reith, R. Podloucky, E.-W. Scheidt, W. Wolf, and M. Marsman, Phys. Rev. B {\bf80}, 064504 (2009).
\bibitem{21} K. Togano, P. Badica, Y. Nakamori, S. Orimo, H. Takeya, and K. Hirata, Phys. Rev. Lett. {\bf93}, 247004 (2004).
\bibitem{22} P. Badica, T. Kondo, and K. Togano, J. Phys. Soc. Jpn. {\bf74}, 1014(2005).
\bibitem{23} E. Bauer, G. Hilscher, H. Michor, C. Paul, E. W. Scheidt, A. Gribanov, Y. Seropegin, H. Noel, M. Sigrist, and P. Rogl,  Phys. Rev. Lett. {\bf92}, 027003 (2004).
\bibitem{24} T. Takayama, K. Kuwano, D. Hirai, Y. Katsura, A. Yamamoto, and H. Takagi, Phys. Rev. Lett. {\bf108}, 237001 (2012).
\bibitem{25} A. Subedi, L. Ortenzi, and L. Boeri, Phys. Rev. B {\bf87}, 144504 (2013).
\bibitem{26} H. Chen, X. Xu, C. Cao, and J. Dai, Phys. Rev. B {\bf86}, 125116 (2012).
\bibitem{27} B. I. Jawdat, B. Lv, X. Y. Zhu, Y. Y.  Xue, and C. W. Chu, Phys. Rev. B {\bf91}, 094514 (2015).
\bibitem{28} K. Hu, B. Gao, Q. Ji, Y. Ma, W. Li, X. Xu, H. Zhang, G. Mu, F. Huang, C. Cai, X. Xie, and M. Jiang, Phys. Rev. B {\bf93} 214510 (2016).
\bibitem{29} K. Hu, B. Gao, Q. Ji, Y. Ma, H. Zhang, G. Mu, F. Huang, C. Cai, and X. Xie, Front. Phys. , {\bf11}, 117403 (2016)
\bibitem{30} B. Lv. B. I. Jawdat, Z. Wu, M. Sorolla, M. Gooch, K. Zhao, L. Deng, Y. Y. Xue, B. Lorenz. A. M. Guloy, and C. W. Chu, Inorg. Chem, {\bf54}, 1049(2015).
\bibitem{31} T. F. Smith, C. W. Chu, and M. B. Maple, Cryogenics {\bf9}, 53 (1969).
\bibitem{32} C. W. Chu, Phys. Rev. Lett. {\bf33}, 1283 (1974).
\bibitem{33} A. Tari, "The Specific Heat of Master at Low Temperatures", Imperial College Press, London, p. {\bf150}, (2003).
\bibitem{34} B. M. Jawdat, B. Lv, Z. Wu, and C. W. Chu, unpublished data.
\bibitem{35} K. Kadowaki, and S. B. Woods, Solid State Commun. {\bf58}, 507-509 (1986).
\bibitem{36} A. C. Jacko, J. O. Fj\ae restad, B. J. Powell Nat. Phys. {\bf5}, 422-425 (2009).
\bibitem{37} T. Smith, Physics Letters {\bf33}A, 465 (1970).
\end{thebibliography}
\end{document}